\documentclass[aps,prl,twocolumn,superscriptaddress,nofootinbib]{revtex4-2}
\usepackage[english]{babel}
\usepackage{bm}
\usepackage{amsmath}
\usepackage{amssymb}
\usepackage{dutchcal}
\usepackage[dvips]{graphicx}
\usepackage[colorlinks=true, allcolors=magenta]{hyperref}
\usepackage{diagbox}
\usepackage{multirow}
\usepackage{xfrac}
\usepackage{xcolor}

\begin{document}
\title{Interplay of interlayer distance and in-plane lattice relaxations in encapsulated twisted bilayers}

\author{V.~V.~Enaldiev}
\email{vova.enaldiev@gmail.com}
\affiliation{Moscow Center for Advanced Studies, Kulakova str. 20, Moscow 123592, Russia} 
\affiliation{Kotelnikov Institute of Radio-engineering and Electronics of the RAS, Mokhovaya 11-7, Moscow 125009, Russia}

\begin{abstract}
	Encapsulation protects functional layers, ensuring structural stability and improving the quality of assembled van der Waals heterostructures. Here, we develop a model that describes lattice relaxation in twisted bilayers accounting for encapsulation effects, incorporated via a single parameter characterizing rigidity of encapsulation material interfaces. By analysing the twist-angle dependence of weak-to-strong lattice relaxation transition in twisted transition metal dichalcogenide bilayers, we show that increasing interface rigidity raises the crossover twist angle between the two relaxation regimes. Furthermore, tuning this rigidity parameter allows to achieve a good agreement with existing experimental results.   
\end{abstract}

\maketitle

\date{\today}

\begin{figure}[t]
	\includegraphics[width=1.0\columnwidth]{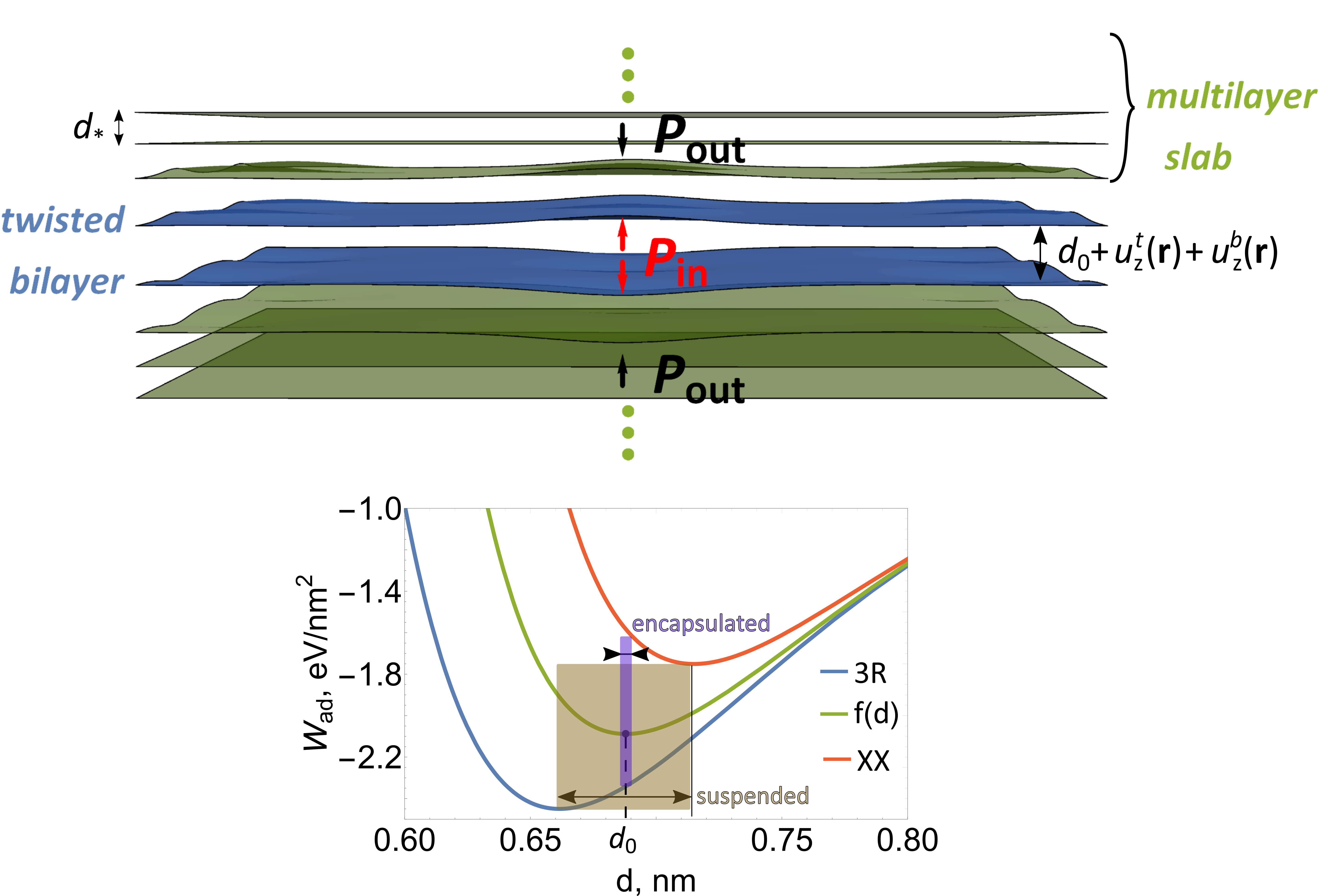}
	\caption{\label{fig:1} (Top) Sketch of interlayer distance relaxation for encapsulated twisted bilayers. Modulation of interlayer distances at twisted interface is determined by balance between stacking-dependent pressure, $P_{\rm in}$, produced by the twisted layers, and pressure, $P_{\rm out}$, arising due to modulation of interlayer distances of interface layers in encapsulating multilayer slabs. (Bottom) Interlayer distance dependence of adhesion energy densities \eqref{Eq:W} for rhombohedral (3R; $\bm{r}_0=(0,a/\sqrt{3})$) and XX ($\bm{r}_0=0$) stackings as well as stacking-averaged values, $\langle W(\bm{r}_0,d)\rangle_{\bm{r}_0}=f(d)$. Semi-transparent boxes schematically show intervals of the adhesion energy densities (height) and interlayer distances (width) set by interlayer distance relaxation in suspended \eqref{Eq:dist_free} (brown) and encapsulated \eqref{Eq:dist_enc} (magenta) twisted bilayers.}
\end{figure}

Van der Waals heterostructures with interlayer twist, such as twisted bi- and tri- layers of graphene \cite{cao2018unconventional,cao2018unconventional,Yankowitz2019,Kazmierczak2021,barrier2024,Craig2024,Pierce2025,Tanaka2025} and transition metal dichalcogenides \cite{Molino2023,Ko2023,Yang2023,Gao2024,VanWinkle2024,Li2024,Ding2024,Zhang2025}, are at the forefront of quantum material research. These twistronic structures exhibit a mesoscale periodic modulation of the interlayer stacking register at the interface, known as moir\'e superlattice, which significantly modifies electronic and optoelectronic properties of constituent layers. 

The formation of moir\'e superlattices is inevitably accompanied with emergence of metastable stacking regions, driving atomic relaxation at the interface to minimize the structure's total energy \cite{AldenPNAS,yoo2019atomic,Weston2020,rosenberger2020}. This results in transformation of moir\'e patterns, specific for rigid layers, into arrays of commensurate domains, characterised by the most energetically favourable stacking and separated by network of strain-absorbing domain walls \cite{Weston2020,Enaldiev_PRL,Engelke2023}. For twisted transition metal dichalcogenide (TMD) layers near parallel (P) alignment, the domains adopt rhombohedral stacking (where metals in one layer align vertically with chalcogens in the neighboring layer), manifesting interfacial ferroelectricity \cite{li2017,Ferreira2021,Weston2022,Wang2022}. In contrast, for anti-parallel (AP) twisted TMD bilayers, the domains adopt inversion symmetric 2H stacking arrangement characterised by vertical alignment of metals with chalcogens and, simultaneously, chalcogens with metals in adjacent layers.
 
A crossover from moir\'e pattern of rigidly twisted layers to the domain structures usually happens at small twist angles where gain in formation of the lowest energy domains overcomes the elastic energy costs required for creation of the domain wall networks. The crossover twist angles $\theta_*\approx 2.5^{\circ}$ ($\theta_*\approx1^{\circ}$) for P (AP) twisted TMD bilayers have been established in Refs. \cite{Weston2020,Enaldiev_PRL}. Although not explicitly stated, these values correspond to the case of suspended bilayers as no encapsulation have been taken into account in the model. However, recent experiments \cite{Kinoshita2024} demonstrate that encapsulation of TMD bilayers in hexagonal borone nitride (hBN) enhances lattice relaxation for the same twist angles compared to suspended bilayers, shifting the crossover twist angles to higher values. 

In this work, we develop a model for lattice relaxation in twisted TMD bilayers that accounts for encapsulation effects. By introducing a lattice relaxation strength parameter, we determine the crossover twist angles characterising the transition between weak and strong relaxation regimes in moir\'e superlattices of encapsulated and suspended twisted bilayers. We demonstrate that suppression of interlayer distance modulation due to encapsulation of twisted bilayers leads to an increase of the crossover twist angle. Moreover, by adjusting a single model parameter, that describes rigidity of twisted bilayer/encapsulation crystal interface, we achieve a quantitative agreement with experimental results \cite{Kinoshita2024}.

Stacking-dependent adhesion energy of the twisted layers is a fundamental property that characterises energetics of interlayer coupling in a moir\'e superlattice. For MX$_2$ bilayers (M=Mo,W; X=S,Se) the interlayer adhesion energy density reads as follows \cite{Weston2020,Enaldiev_PRL}:
\begin{multline}\label{Eq:W}
	W_{\rm ad}(\bm{r}_0,d) = f(d) + \\
	 \sum_{l=1,2,3}\left[Ae^{-qd}\cos\left(\bm{G}_l\bm{r}_0\right) +Be^{-Gd}\sin\left(\bm{G}_l\bm{r}_0+\phi\right)\right]. 
\end{multline}
Here, $\bm{r}_0$ is an in-plane offset between aligned TMD layers with the origin, $\bm{r}_0=(0,0)$, at XX stacking, $d$ is an interlayer distance, $\bm{G}_{1,2,3}$ is triad of the shortest reciprocal vectors of a single layer, related by $\pm120^{\circ}$-rotation ($\left|\bm{G}_{1,2,3}\right|=G=4\pi/a\sqrt{3}$, $\bm{G}_{2,3}=\hat{R}^{\pm1}_{3}\bm{G}_{1}$); $A$, $B$, $Q$ and $\varepsilon$ are material-dependent fitting parameters (see \cite{Enaldiev_PRL}), $\phi=\pi/2$ and $\phi=0$ for P and AP alignment of layers, respectively. 

Lattice relaxation in twisted TMD bilayers involves not only moir\'e pattern reconstruction but also modulation of interlayer distances. A common approximation \cite{Enaldiev_PRL,Zhou2015,CarrPRB2018} assumes that the interlayer distance $d(\bm{r}_0)$ locally optimizes the adhesion energy, corresponding to vanishing normal pressure for every $\bm{r}_0$:
\begin{equation}\label{Eq:Pin}
	P_{\rm in}(\bm{r}_0,d) = -\frac{\partial W_{\rm ad}(\bm{r}_0,d)}{\partial d}=0.
\end{equation} 
Near the energy extrema (Fig. \ref{fig:1}), the interlayer adhesion energy \eqref{Eq:W} is dominated by $f(d)$ term, so that the interlayer distances vary in vicinity of the optimal value, $d_0$, where $f(d)\approx f(d_0)+\varepsilon(d-d_0)^2$ \cite{Enaldiev_PRL}. By expanding the exponential prefactors in Eq. \eqref{Eq:W} to the first order in $d-d_0$ and substituting into Eq. \eqref{Eq:Pin}, we obtain the stacking-dependent distance modulation \cite{Enaldiev_PRL}:
\begin{equation}\label{Eq:dist_free}
 	d_{\rm f}(\bm{r}_0)=d_0+\frac{1}{2\varepsilon}\sum_{l=1,2,3}\left[\mathcal{A}q\cos\left(\bm{G}_l\bm{r}_0\right)+\mathcal{B}G\sin\left(\bm{G}_l\bm{r}_0+\phi\right)\right],
\end{equation}
where $\mathcal{A}\equiv Ae^{-qd_0}$ and $\mathcal{B}\equiv Be^{-Gd_0}$. Strictly speaking, Eq. \eqref{Eq:dist_free} and similar dependences in Refs. \cite{Zhou2015,CarrPRB2018} describe suspended twisted TMD bilayers, as the models do not comprise any information about property of the interface with encapsulation material.

To take into account encapsulation effects, we model twisted MX$_2$ bilayers as embedded between multilayer slabs having an equilibrium stacking characterised by interlayer distance $d_*$ in the bulk (see Fig. \ref{fig:1}). We suppose that in the slabs an areal density of the adhesion energy between the layers depends only on distance between atomic planes and for small deviations ($|d-d_*|\ll d_*$) can be approximated by a harmonic potential:
\begin{equation}\label{Eq:enc_adh}
	W_{u/d}(d) = \frac{1}{2}k_{u/d}(d-d_*)^2,
\end{equation}
where $k_{u/d}$ is an effective parameter characterizing interlayer binding in the up (u)/down (d) slab. In bulk slabs with a unit cell comprising two atomic planes (like hBN), the parameter $k$ would determine a frequency of zero quasi-momentum optical phonon, $\omega_{\rm BM}^2=4k/\rho$, ($\rho$ is areal mass density for a single layer of a slab), characterised by out-of-plane polarisation of displacements. However, for the slab layers at the interface with the TMD bilayer, this parameter may differ from its bulk value, for example, due to lattice incommensurability. Therefore, below we treat $k_{u,d}=k$ as phenomenological parameter describing the bilayer/slab interfaces and examine how the lattice relaxation depends on its magnitude. 

To describe interlayer distance relaxation for the encapsulated bilayers we introduce out-of-plane displacements $u_z^{t/b}(\bm{r}_0)$ of the top/bottom layer with respect to stacking-averaged interlayer distance $d_0$. For each local lateral offset $\bm{r}_0$ in moir\'e supercell, the TMD bilayer produces pressure $P_{\rm in}(\bm{r}_0, d_0+u_z^t+u_z^b)$ at the interface layer of the u/d slab, which, at the equilibrium, should be balanced by the pressure, $P_{\rm out}(d_*+u_z^{t/b})=-\partial W_{\rm u/d}/\partial d|_{d=d_*+u_z^{t/b}}$ arising due to shift of the atomic plane by $u_z^{t/b}$ in the u/d slab. Equalizing $P_{\rm in}$ and $P_{\rm out}$ at the two interfaces we obtain the following modulation of interlayer distances in the TMD bilayer:
\begin{eqnarray}\label{Eq:dist_enc}
	d_{\rm enc}(\bm{r}_0)\equiv d_0+u_z^t(\bm{r}_0)+u_z^b(\bm{r}_0)=\qquad\qquad\qquad\qquad\qquad\\
	d_0+\frac{\sum_{l=1,2,3}\left[\mathcal{A}q\cos\left(\bm{G}_l\bm{r}_0\right)+\mathcal{B}G\sin\left(\bm{G}_l\bm{r}_0+\phi\right)\right]}{2\varepsilon+\frac{k_{\rm u}k_{\rm d}}{k_{\rm u}+k_{\rm d}}}. \nonumber
\end{eqnarray}
For derivation of Eq. \eqref{Eq:dist_enc} we applied the same approximation for the $d$-dependent factors in $W_{\rm ad}$ as in the case of non-encapsulated bilayers \eqref{Eq:dist_free}. In the absence of encapsulation slabs, $k_{\rm u}=k_{\rm d}=0$, Eq. \eqref{Eq:dist_enc} reduces to the result for suspended bilayers \eqref{Eq:dist_free}, whereas in opposite limit of rigid slabs, $k_{\rm u},k_{\rm d}\gg4\varepsilon$, the stacking variation of the interlayer distances is completely suppressed with $d(\bm{r}_0)\equiv d_0$. So, encapsulation primarily influences the range of accessible interlayer distances, which governs the energetics of interlayer adhesion in the moiré superlattice via substitution $d\to d_{\rm enc}(\bm{r}_0)$ in Eq. \eqref{Eq:W}.



Having incorporated the characteristics of encapsulation slabs in the model for adhesion energy we study how the encapsulation affects the in-plane relaxation of moir\'e pattern in twisted TMD bilayers. The lattice relaxation \cite{Weston2020,Enaldiev_PRL} is described by in-plane displacement fields $\bm{u}_{t}(\bm{r})$ and $\bm{u}_{b}(\bm{r})$ in top and bottom layers of the bilayer, that are determined by minimization of sum of in-plane elastic and the local adhesion energy densities of the bilayers over a supercell (sc),
\begin{multline}\label{Eq:inplaneE}
	\mathcal{E}=\int\limits_{\rm sc}d^2\bm{r}\left\{\sum_{l=t,b}\left[(\lambda_{l}/2)u^{(l)}_{ii}u^{(l)}_{ii}+ \mu_l u^{(l)}_{ij}u^{(l)}_{ji}\right]+\right.\\
	\left. +W_{\rm ad}\left[\bm{r}_0(\bm{r}),d_{\rm enc}(\bm{r}_0(\bm{r}))\right]\right\}.
\end{multline}
Here $\lambda_{t/b}$ and $\mu_{t/b}$ are Lam\'e parameters, characterising in-plane rigidity of the top/bottom layer, $2u^{t/b}_{ij}=\partial_iu^{t/b}_j+\partial_ju^{t/b}_i$ is in-plane strain tensor ($i,j=\{x,y\}$), 
\begin{equation}\label{Eq:ord_param}
	\bm{r}_0(\bm{r})=\theta\hat{z}\times\bm{r}+\bm{u}_t-\bm{u}_b
\end{equation} 
is the in-plane interlayer lateral offset in the bilayer with interlayer twist angle $\theta$, that provides moir\'e superlattice periodicity of the adhesion energy density. To find minimum of the energy functional with periodicity of moir\'e superlattice we use Fourier series expansion $\bm{u}_{t/b}(\bm{r})=\sum_{n}\widetilde{\bm{u}}^{(n)}_{t/b}e^{i\bm{g}_n\bm{r}}$ over reciprocal vectors of moir\'e superlattice $\bm{g}_{n}=\theta \bm{G}_n\times z$, leaving for every superlattice period enough harmonics to attain convergence.  
The Fourier amplitudes, $\widetilde{\bm{u}}_{t/b}^{(n)}$, are numerically determined from miminization of the energy functional \eqref{Eq:inplaneE}. 

\begin{figure}
	\includegraphics[width=\columnwidth]{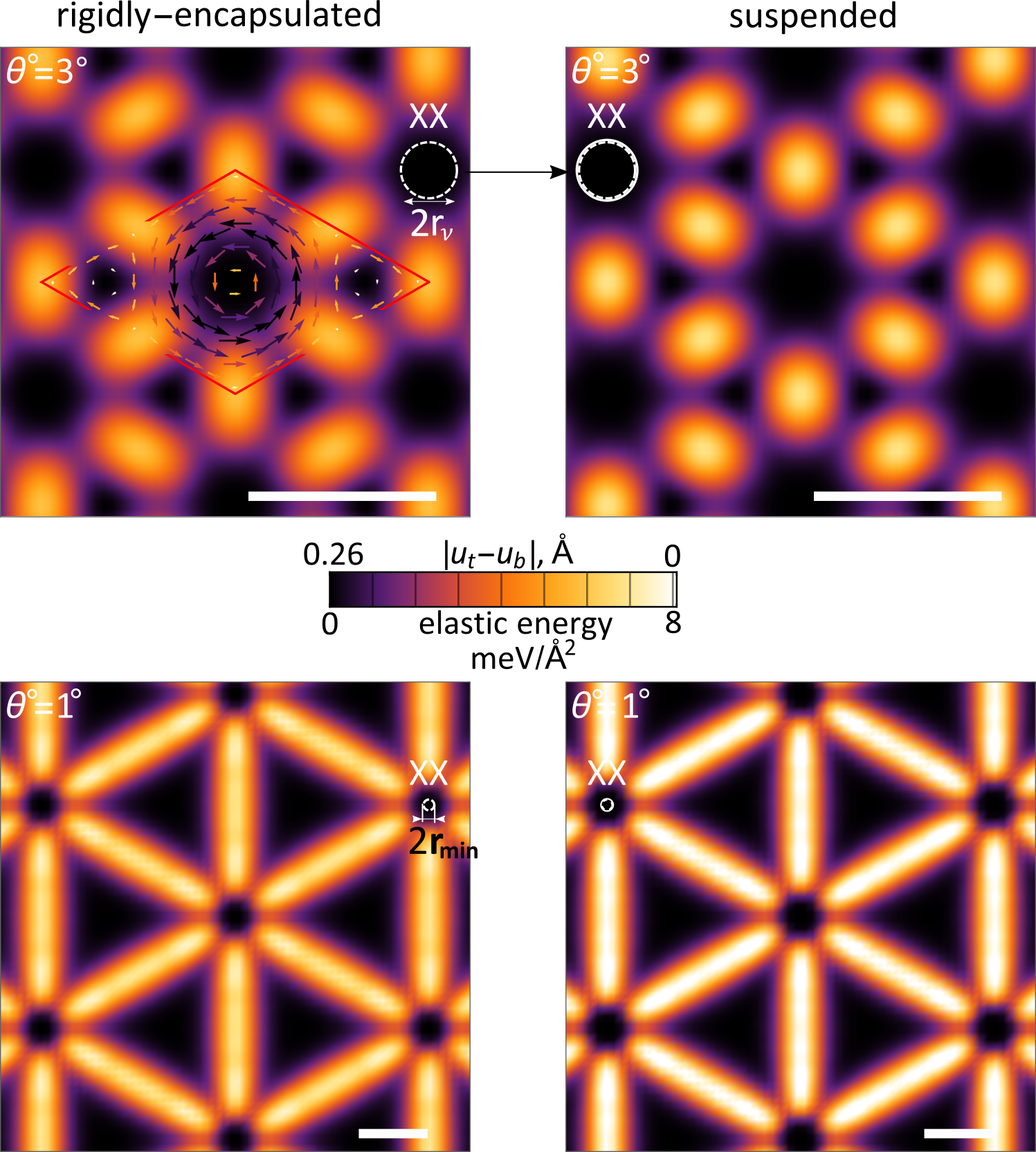}
	\caption{\label{fig:2} Densities of elastic energy for rigidly-encapsulated ($k_{u,d}\gg 4\varepsilon$) and suspended  ($k_{u,d}=0$) lattice-relaxed P WSe$2$ bilayers at $\theta^{\circ}=3^{\circ}$ and $\theta^{\circ}=1^{\circ}$. Circles highlight XX-stacked regions superimposed on the right panels for comparison. At marginal twist angles, $\theta^{\circ}=1^{\circ}$, radius, $r_{\rm min}$, of the XX stacking areas becomes independent on the twist angle. Inset on the top left panel demonstrates distribution of relative displacement field, $\bm{u}_t-\bm{u}_b$, throughout a moir\'e supercell. Scale bars are 5 nm.}
\end{figure}

We demonstrate the results of calculations for P WSe$_2$ bilayers ($\phi=0$ in Eqs. \eqref{Eq:W},\eqref{Eq:dist_enc}), where experimental studies have systematically investigated the twist-angle dependence of lattice relaxation strength in hBN-encapsulated structures \cite{Kinoshita2024}. In Fig. \ref{fig:2} we show maps of the in-plane elastic energy densities for lattice-relaxed P WSe$_2$ bilayers encapsulated in perfectly rigid ($k_{u/d}\gg4\varepsilon$) and suspended ($k_{u/d}=0$) slabs, with interlayer misalignments $\theta^{\circ}=3^{\circ}$ and $\theta^{\circ}=1^{\circ}$. 

At $\theta^{\circ}=3^{\circ}$, relaxation of moir\'e superlattice is characterised by the expansion of energetically favourable rhombohedral-stacked (3R) regions and contraction of the highest energy XX-stacked regions. The  key difference between the relaxed moir\'e patterns of suspended and rigidly-encapsulated bilayers lies in the smaller size of XX-stacked regions in the latter, resulting from the reduced range of interlayer distance variation (see Fig. \ref{fig:1}). The corresponding relative displacement field, $\bm{u}_t-\bm{u}_b$, exhibits clockwise and anti-clockwise rotations around the centers of rhombohedral and XX stacking areas, respectively (top left panel in Fig. \ref{fig:2}). As the twist angle decreases, the 3R-stacked regions evolve into triangular domains, separated by domain walls that intersect at areas of XX stacking. Notably, the size of these XX-stacked regions becomes independent of $\theta$ at marginal twists (see below).  

Following Ref. \cite{Kinoshita2024} we quantify the strength of lattice relaxation in twisted bilayers using the ratio,
\begin{equation}\label{Eq:relax_param}
	R_{\nu}(\theta)=\frac{S^{\rm XX}_{\nu}(\theta)}{S_{\rm sc}(\theta)},
\end{equation} 
where $S_{\rm sc}(\theta)=(a/\theta)^2\sqrt{3}/2$ is the area of a moir\'e supercell, and $S^{\rm XX}_{\nu}(\theta)$ is an area of relaxed moir\'e superlattice corresponding to XX stacking of the layers. Specifically, we define $S^{\rm XX}_{\nu}(\theta)$ as a circle of radius $r_{\nu}$ centered on the XX-stacked region, within which the deviations of $\bm{r}_0(\bm{r})$ \eqref{Eq:ord_param} from its value at the center are less than a small fraction, $\nu$ ($\nu\ll 1$), of the monolayer lattice constant\footnote{Strictly speaking, according to the definition of $\bm{r}_0$ (see Eq. \eqref{Eq:W}), $S^{\rm XX}$ should correspond to an area where $\bm{r}_0(\bm{r})=n\bm{a}_1+m\bm{a}_2$ ($\bm{a}_{1,2}$ are basis vectors of a single-layer Brave lattice) for a moir\'e supercell indexed by integers $\{n,m\}$. However, this condition is too strong and is satisfied only at a single point -- the middle of the XX-stacked regions -- resulting in $S^{\rm XX}=0$. This occurs because for every $\theta$ the relative displacements field $\bm{u}^t-\bm{u}^b$, which describes rotation around the XX-stacked regions (see upper-left panel in Fig. \ref{fig:2}), does not fully cancel the geometric interlayer rotation $\theta z\times\bm{r}$ for $\bm{r}_0(\bm{r})$ \eqref{Eq:ord_param}.}. We checked that the exact choice of $\nu$ does not affect the conclusions presented below provided $\nu\ll1$.

\begin{figure}[!t]
	\includegraphics[width=\columnwidth]{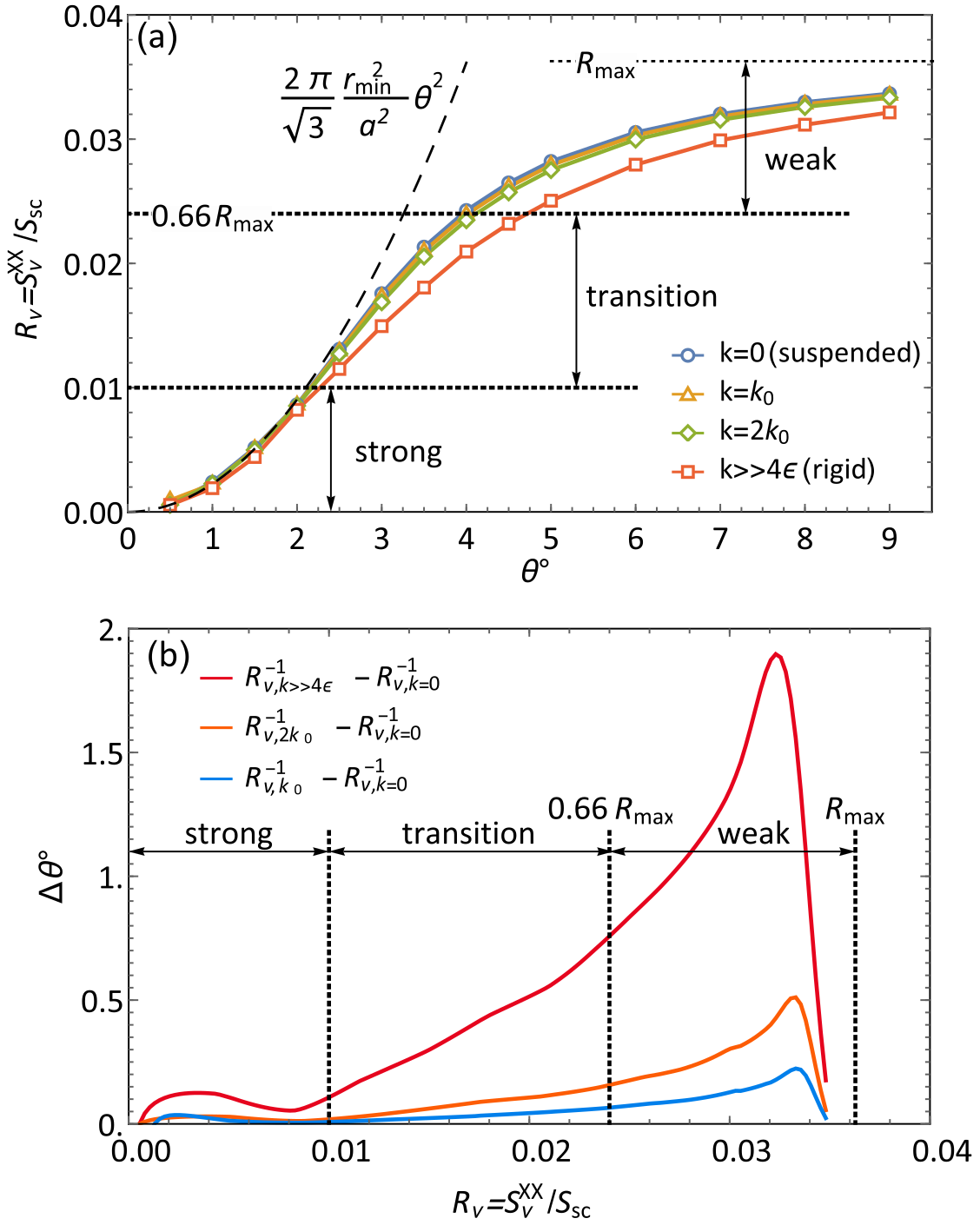}
	\caption{(a) Twist-angle dependences of the relaxation strength parameter $R_{\nu}(\theta)$ \eqref{Eq:relax_param} computed with $\nu=0.1$ for four values of the parameter $k$ describing rigidity of P WSe$_2$ bilayer/slab interfaces. Value $0.66R_{\rm max}$ sets up the crossover twist angles for considered encapsulation cases. (b) Difference between the inverse dependences, shown in (a), for encapsulated and suspended P WSe$_2$ bilayers. In case of rigid bilayer/slab interface the crossover twist angle is increase by $\approx 0.75^{\circ}$ compared to the suspended bilayers in a good agreement with experimental results \cite{Kinoshita2024}.\label{fig:3}}
\end{figure}

Figure \ref{fig:3}(a) presents the dependences of $R_{\nu}(\theta)$ \eqref{Eq:relax_param} for P WSe$_2$ bilayers with interlayer twist angles $0^{\circ}<\theta^{\circ}\leq 9^{\circ}$, computed using $\nu=0.1$. We examine four cases of the phenomenological parameter $k_{u}=k_{d}=k$\footnote{The value $k_0$ is estimated from the areal single-layer density $\rho$ and the out-of-plane $\Gamma$-point optical phonon frequency $\omega_{\rm BM}\approx15$\,meV in the bulk hBN \cite{Serrano2007}.}: (i) $k=0$ (ii) $k=k_0\equiv\omega_{\rm BM}^2\rho/4\approx 317$\,eV/nm$^4$, (iii) $k=2k_0$, and (iv) $k\gg 4\varepsilon=760$\,eV/nm$^4$ \cite{Enaldiev_PRL}. Here, case (i) represents a freely suspended WSe$_2$ bilayer, while case (iv) corresponds to perfectly rigid bilayer/slab interface. Case (ii) describes the interlayer binding strength in bulk hBN, and case (iii) models enhanced coupling between the interface layer of the slab and the WSe$_2$ bilayer or the other slab layers.

For large twist angles, $\theta^{\circ}\gtrsim5^{\circ}$, $R_{\nu}(\theta)$ approaches to the $k$- and $\theta$- independent limit, $R_{\rm max}=2\pi \nu^2/\sqrt{3}$, determined by the area of XX stacking in the non-reconstructed moir\'e superlattice, $S^{\rm XX}_{\nu}=\pi (\nu a/\theta)^2$. In this interval of twist angles, lattice relaxation is weak due to exceedingly high elastic energy costs compared to gains in the adhesion energy, resulting in vanishingly small displacement fields. 

In contrast, for small twist angles ($\theta^{\circ}<2^{\circ}$), strong relaxation of the moir\'e pattern results in shrinkage of XX-stacked regions into the domain wall network nodes, where they are characterised by a twist-angle-independent area, $S^{\rm min}_{\rm XX}=\pi r^2_{\rm min,\nu}$, with $r_{\rm min,\nu}$ determined by the adhesion energy density $W_{\rm ad}(\bm{r}_0,d_{\rm enc}(\bm{r}_0))$ evaluated at $|\bm{r}_0|/a=\nu$. Consequently, in this regime, $R_{\nu}(\theta)$ follows $\theta^2$-scaling: 
\begin{equation}\label{Eq:R_theta2}
	R_{\nu}(\theta)\approx\frac{2\pi}{\sqrt{3}}\frac{r^2_{\rm min,\nu}}{a^2}\theta^2,
\end{equation} plotted by the dashed line in Fig. \ref{fig:3}(a). This behaviour becomes universal for all encapsulation cases in the strong relaxation regime because the XX-stacked region occupies only a minor fraction of the moir\'e supercell (see circles on the bottom panels in Fig. \ref{fig:2}) and $r_{\rm min,\nu}$ is of weak dependence on the encapsulation strength $k$.  

As the twist angle decreases, moir\'e superlattices passes from weak to strong relaxation regimes over an interval of twist angles specifying transition from the two asymptotic behaviours of $R_{\nu}(\theta)$ (Fig. \ref{fig:3}(a)). Although the transition is continuous in $\theta$, we define the crossover twist angle, $\theta_*$, as the value corresponding to the upper boundary of transition interval $R_{\nu}(\theta_*)\approx 0.66R_{\rm max}$ characterising change of the slope for  $R_{\nu}(\theta)$-dependences. In the case of suspended bilayers the crossover twist angle is $\theta_*\approx 3.8^{\circ}$, whereas with rise of the rigidity of bilayer/slab interface it gradually increases attaining $\theta_*\approx4.5^{\circ}$ for perfectly rigid case (iv). This results from the increasing energy difference between XX and 3R stackings brought about by contraction of the interlayer distance modulation range (Fig. \ref{fig:1}).   

To explicitly illustrate the difference of $\theta_*$ between encapsulated and suspended bilayers, Fig. \ref{fig:3}(b) shows $R_{\nu,k}^{-1}-R_{\nu,k=0}^{-1}$ across the full range of $R$ for $k$ values representing the three encapsulation cases. The largest difference $\Delta \theta_*\approx 0.75^{\circ}$ occurs at the crossover point $0.66R_{\rm max}$ for encapsulation with a perfectly rigid bilayer/slab interface (i.e. at $k\gg 4\varepsilon$). The experimentally measured difference \cite{Kinoshita2024} of $\Delta\theta_*\approx 1^{\circ}$ between hBN-encapsulated and suspended P WSe$_2$ bilayers agrees well with the perfectly rigid interface model, likely due to incommensurability between TMD and hBN lattices suppressing interlayer distance modulation. 

To summarize, we have developed a model that accounts for encapsulation effects on lattice relaxation of moir\'e pattern in twisted TMD bilayers. The model incorporates a single phenomenological parameter characterising rigidity of each of the two bilayer/encapsulation crystal interfaces. We demonstrate that increased interface rigidity noticeably raises the crossover twist angles separating weak and strong moir\'e superlattice relaxation regimes. The affect arises because encapsulation suppresses variation of interlayer distances across the moir\'e superlattice. 

By modeling encapsulated P WSe$_2$ bilayers, we show that assuming perfectly rigid interfaces leads to a good agreement between the predicted crossover twist angle and experimental results \cite{Kinoshita2024} for hBN-encapsulated P WSe$_2$ bilayers. Such a high interface rigidity likely arises due to lattice incommensurability at the TMD bilayer/hBN interfaces.

Finally, we note that the model can be extended to other twistronic heterostructures. For instance, in WX$_2$/MoX$_2$ heterobilayers, encapsulation-induced shrinkage of XX-stacked regions and reduction of interlayer distance variation could considerably change magnitude of strain inside them influencing interlayer exciton binding energies \cite{Soltero2024} and twirling of domain wall network in long-period moir\'e superlattices \cite{Kaliteevski2023,Jong2023,Mesple2023}
 
{\it Acknowledgements.} I thank Vladimir Fal'ko and Anvar Baimuratov for fruitful discussions. The work was supported by the Ministry of Science and Higher
Education of the Russian Federation (Goszadaniye FSMG-2023-0011).


\bibliography{refer}

\end{document}